\documentclass[aps,%
prx,%
twocolumn,%
superscriptaddress,%
longbibliography,%
reprint,%
floatfix%
]{revtex4-2}

\usepackage[english]{babel}
\usepackage{graphicx}
\usepackage{amsmath,amssymb}
\usepackage{bm}
\usepackage{mathrsfs}
\usepackage{xcolor}
\usepackage{booktabs}
\usepackage{xpatch}
\usepackage{natbib}

\graphicspath{{figures/}}

\makeatletter
\let\ORIbbl@fixname\bbl@fixname
\def\bbl@fixname#1{%
\@ifundefined{languagealias@\expandafter\string#1}
{\ORIbbl@fixname#1}
{\edef\languagename{\@nameuse{languagealias@#1}}}%
}
\newcommand{\definelanguagealias}[2]{%
\@namedef{languagealias@#1}{#2}%
}
\makeatother

\definelanguagealias{en}{english}
\definelanguagealias{en-US}{english}
\definelanguagealias{zh}{chinese}

\newcommand{\dd}{\,\mathrm{d}}
\newcommand{\mi}{\mathrm{i}}
\newcommand{\ee}{\mathrm{e}}
\newcommand{\gagg}{g_{a\gamma\gamma}}
\newcommand{\cc}{\mathrm{c.c.}}
\newcommand{\arem}{\mathrm{AREM}}
\DeclareMathOperator{\sinc}{sinc}

\begin{document}

\title{Axion generation and detection in laser-plasma wakefields}

\author{Xiangyan An}
\affiliation{State Key Laboratory of Dark Matter Physics, Tsung-Dao Lee Institute, Shanghai Jiao Tong University, Shanghai 201210, China}
\affiliation{State Key Laboratory of Dark Matter Physics, Key Laboratory for Laser Plasmas (MoE), School of Physics and Astronomy, Shanghai Jiao Tong University, Shanghai 200240, China}

\author{Min Chen}
\email{minchen@sjtu.edu.cn}
\affiliation{State Key Laboratory of Dark Matter Physics, Key Laboratory for Laser Plasmas (MoE), School of Physics and Astronomy, Shanghai Jiao Tong University, Shanghai 200240, China}

\author{Jianglai Liu}
\email{jianglai.liu@sjtu.edu.cn}
\affiliation{State Key Laboratory of Dark Matter Physics, Tsung-Dao Lee Institute, Shanghai Jiao Tong University, Shanghai 201210, China}

\author{Zhan Bai}
\affiliation{State Key Laboratory of Ultra-intense Laser Science and Technology, Shanghai Institute of Optics and Fine Mechanics, Chinese Academy of Sciences, Shanghai 201800, China}

\author{Liangliang Ji}
\affiliation{State Key Laboratory of Ultra-intense Laser Science and Technology, Shanghai Institute of Optics and Fine Mechanics, Chinese Academy of Sciences, Shanghai 201800, China}

\author{Zhengming Sheng}
\affiliation{State Key Laboratory of Dark Matter Physics, Tsung-Dao Lee Institute, Shanghai Jiao Tong University, Shanghai 201210, China}
\affiliation{State Key Laboratory of Dark Matter Physics, Key Laboratory for Laser Plasmas (MoE), School of Physics and Astronomy, Shanghai Jiao Tong University, Shanghai 200240, China}

\author{Jie Zhang}
\affiliation{State Key Laboratory of Dark Matter Physics, Tsung-Dao Lee Institute, Shanghai Jiao Tong University, Shanghai 201210, China}
\affiliation{State Key Laboratory of Dark Matter Physics, Key Laboratory for Laser Plasmas (MoE), School of Physics and Astronomy, Shanghai Jiao Tong University, Shanghai 200240, China}

\date{\today}

\begin{abstract}
    The axions are compelling candidates for cold dark matter, but their extremely weak interaction with photons makes laboratory searches challenging.
    We show that the quasi-static electromagnetic fields of a laser-plasma wakefield, which can exceed $10^{11}$\,V/m, enable axion generation without an external production magnet and enhance the conversion rate by two orders of magnitude over a conventional magnetic production region.
    Self-consistent particle-in-cell simulations reveal two complementary routes to detection.
    In the first route, axions are reconverted into photons within the wakefield and laser fields, eliminating the need for a separate regeneration magnet but requiring to suppress the intense laser-plasma background. The regenerated photons have polarization, harmonic-frequency, and Laguerre-Gaussian transverse-mode signatures that are largely absent from the driving fields, allowing successive filters to isolate the signal.
    In the second route, axions traverse a wall and undergo reconversion in a downstream magnet, providing a much lower background at the cost of requiring both the magnet and a seed pulse for coherent amplification.
    For axion masses below $0.1$\,meV, meter-scale wakefield guiding under our stated assumptions yields a projected coupling sensitivity down to $3.9\times10^{-12}\,\mathrm{GeV}^{-1}$, surpassing the projected constraint of next-generation laboratory searches.
    These results establish ultra-strong plasma wakefields as a magnet-free axion source with two experimentally distinct and complementary detection strategies.
\end{abstract}

\maketitle

\section{Introduction}
\label{sec:introduction}
The axion is a hypothetical pseudoscalar particle originally introduced to resolve the strong CP problem in quantum chromodynamics (QCD)~\cite{peccei_cp_1977,peccei_constraints_1977,wilczek_problem_1978,weinberg_new_1978,kim_axions_2010}.
More broadly, axion-like particles (ALPs) provide a natural extension to the Standard Model (SM), with a two-photon coupling $\gagg$ and axion mass $m_a$~\cite{svrcek_axions_2006,Gelmini:1980re,Bellazzini:2017neg,ema_flaxion_2017,Calibbi:2016hwq,Irastorza:2018dyq}.
Beyond solving fundamental particle physics puzzles, axions are also considered a compelling candidate for cold dark matter, potentially permeating the entire universe~\cite{Preskill:1982cy,Abbott:1982af,Dine:1982ah,Marsh:2015xka}. This has motivated extensive experimental programs aimed at detecting their extremely weak interactions with electromagnetic fields.

Conventional axion searches rely on electromagnetic coupling effects. Haloscope experiments exploit resonant cavities to convert galactic axions into microwave photons under strong magnetic fields~\cite{Sikivie:1983ip,bartram_search_2021,CAPP:2020utb,Ouellet:2018beu}, while helioscopes aim to detect solar axions via X-ray production in magnetic setups~\cite{cast_collaboration_new_2017,armengaud_physics_2019}.
Laboratory-based approaches, such as ``light-shining-through-a-wall'' (LSW) experiments, including Any Light Particle Search II (ALPS-II)~\cite{ehret_new_2010,bahre_any_2013,chou_search_2008,afanasev_experimental_2008,pugnat_results_2008,robilliard_no_2007}, employ lasers and magnetic fields to generate axions, which subsequently traverse a barrier and reconvert into detectable photons.
Polarization-based methods, including vacuum birefringence measurements, further probe axion-induced modifications to photon propagation~\cite{ejlli_pvlas_2020,battesti_bmv_2008,cameron_search_1993,chen_q_2007}.
Besides, the interaction between X-ray free electron lasers and crystals is also used to detect the axions~\cite{halliday_bounds_2025}.
However, these techniques demand both extremely high laser and magnetic fields to achieve detectable signals due to the extraordinarily weak coupling between axions and photons.

Recent advances in laser and plasma wakefield accelerators~\cite{blumenfeld_energy_2007,wang_free-electron_2021,pompili_free-electron_2022,zhu_experimental_2023,picksley_matched_2024} offer a promising new avenue for axion generation and detection.
The wakefields driven by either lasers or particle beams can sustain quasi-static electric fields exceeding $10^{11}$\,V/m~\cite{esarey_physics_2009,mangles_monoenergetic_2004,geddes_high-quality_2004,faure_laserplasma_2004}, which are two orders of magnitude stronger than the equivalent fields produced by state-of-the-art magnets.
These fields arise from charge separation in plasma, providing a robust environment for axion production via the Primakoff effect.
 In parallel, plasma channels enable guided propagation of intense laser pulses over centimeter-to-meter scales~\cite{miao_optical_2020,picksley_matched_2024,miao_benchmarking_2024,shrock_guided_2024,zhu_experimental_2023}, effectively overcoming diffraction limitations and substantially extending the interaction length compared to free-space laser propagation.
In addition, the short laser pulse durations (at the femtosecond scale) facilitate a time-gated background-free measurement with the LSW setup.
Collectively, these capabilities have revived interest in axion studies based on laser-plasma interactions~\cite{tercas_axion-plasmon_2018,burton_plasma-based_2018,mendonca_axion_2020,mendonca_axion_2020-1,huang_axion-like_2022}.

\begin{figure}[t]
    \centering
    \includegraphics[width=\columnwidth]{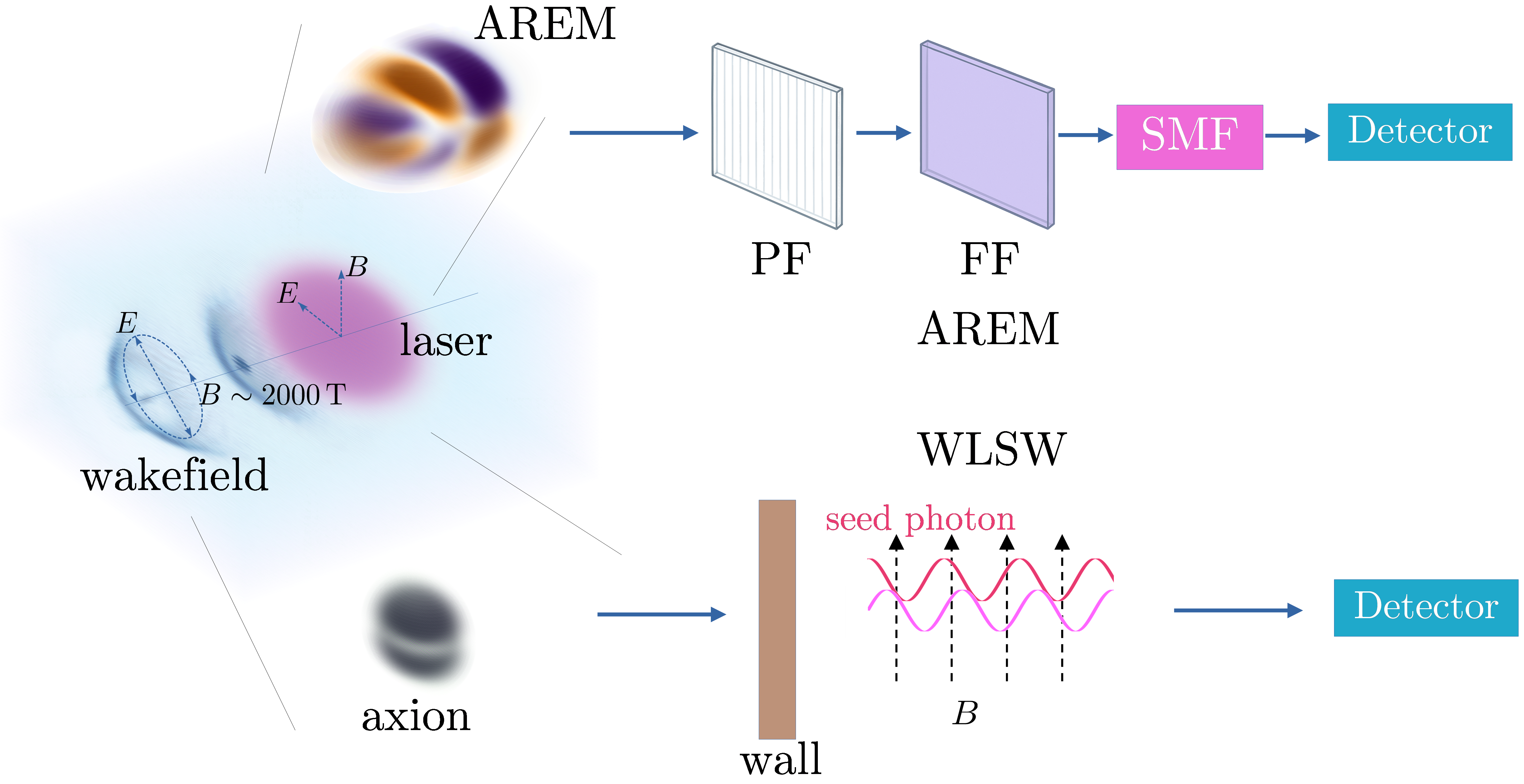}
    \caption{Schematic of the wakefield platform and its two detection routes. An intense laser co-propagates with the plasma wakefield and generates axions. The AREM produced in the same interaction region can be selected using the polarization filter (PF), frequency filter (FF), and spatial-mode filter (SMF), or the axions can pass through a wall into a regeneration magnet.}
    \label{fig:overview}
\end{figure}

This work presents a first-of-its-kind scheme for in situ axion generation and conversion in plasma wakefields, achieved through a unified framework that integrates axion electrodynamics into the PIC code EPOCH~\cite{arber_contemporary_2015,an_modeling_2024}.
Unlike earlier plasma-based proposals that primarily rely on the longitudinal wakefield and still require an external static magnet, we present a magnet-free, purely laser-wakefield interaction scheme for axion production.
We demonstrate self-consistently that the strong laser fields in plasma wakefields can drive axion generation via the Primakoff effect, achieving a significantly enhanced production rate and rendering the scheme attractive for implementation in a LSW configuration.
Figure~\ref{fig:overview} outlines the two detection routes examined in this study. In the first route, the axion-regenerated electromagnetic fields (AREM) generated inside the wakefield are isolated using filters that select specific polarization, frequency, and spatial-mode characteristics. In the second route, the axion pulse exits the plasma, passes through a wall, and enters a conventional magnetic regeneration stage. Coherent seeded regeneration for short-pulse axion detection was introduced in Ref.~\cite{an_coherently_2026}. Here we adapt this technique to the wakefield platform and investigate the combined effects of seed-number and phase uncertainties on the detection performance.

This paper is organized as follows. Section~\ref{sec:theory} establishes the theoretical framework, beginning with the axion electrodynamics in the wakefield configuration, followed by the derivation of the axion source term and its coherent growth, and concluding with the characterization of AREM generation, including its frequency structure and transverse-mode signatures. Section~\ref{sec:simulations} describes the numerical implementation and validates these theoretical predictions through self-consistent particle-in-cell simulations. In Sec.~\ref{sec:phase_matching}, we analyze the phase-matching conditions that govern long-distance accumulation and compare the performance of laser-driven and beam-driven wakefield configurations. Sections~\ref{sec:direct_detection} and~\ref{sec:seeded_detection} are devoted to the two complementary detection routes outlined in Fig.~\ref{fig:overview}: Sec.~\ref{sec:direct_detection} examines direct AREM detection via multi-stage polarization, frequency, and spatial-mode filtering within the wakefield, while Sec.~\ref{sec:seeded_detection} investigates the seeded wakefield LSW configuration, including the impact of seed-number and phase fluctuations on the projected sensitivity. Finally, Sec.~\ref{sec:conclusion} summarizes our main findings and discusses prospects for experimental realization.

\section{Theoretical analysis}
\label{sec:theory}

\subsection{Axion electrodynamics and wakefield configuration}

To be convenient, we use natural units $\hbar=c=\varepsilon_0=1$ except when quoting laboratory parameters. The axion-photon system is described by
\begin{equation}
    \mathcal L=-\frac14F_{\mu\nu}F^{\mu\nu}
    +\frac12\partial_\mu\phi\partial^\mu\phi
    -\frac12m_a^2\phi^2
    -\frac{\gagg}{4}\phi F_{\mu\nu}\widetilde F^{\mu\nu}.
\end{equation}
Here $\mathcal L$ is the Lagrangian density, $\phi$ is the axion field, $F_{\mu\nu}$ is the electromagnetic field tensor, and $\widetilde F^{\mu\nu}$ is its dual. $m_a$ and $\gagg$ denote the axion mass and axion-photon coupling, respectively.
We separate the electromagnetic fields into a background, $\bm E_0,\bm B_0$, and the first-order AREM, $\bm E_1,\bm B_1$. The background contains the laser and plasma fields. To leading order in the weak coupling,
\begin{gather}
    (\partial_t^2-\nabla^2+m_a^2)\phi
    =\gagg\bm E_0\cdot\bm B_0,\label{eq:axion_KG}\\
    \nabla\times\bm E_1=-\partial_t\bm B_1,\label{eq:faraday_1}\\
    \nabla\times\bm B_1=\partial_t\bm E_1
    +\gagg\left[(\partial_t\phi)\bm B_0
    -\bm E_0\times\nabla\phi\right].\label{eq:ampere_1}
\end{gather}
Given the weak coupling, we treat $\bm E_1,\bm B_1$ as perturbations and independently~\cite{an_modeling_2024}. The drive laser propagates along $+x$ and is polarized along $y$,
\begin{equation}
    E_{0,y}^{(\gamma)}=\frac{\omega_0}{k_0}B_{0,z}^{(\gamma)}
    =\frac{a_0}{2}\mathcal F_\gamma(\zeta_0,y,z)\ee^{\mi\xi_0}+\cc,
    \label{eq:laser_field}
\end{equation}
where the superscript $(\gamma)$ denotes the laser field, $\omega_0$ and $k_0$ are the laser frequency and wave number, $a_0$ is the normalized peak laser amplitude, $\xi_0=k_0x-\omega_0t$ is the carrier phase, $\zeta_0=x-v_gt$ is the co-moving coordinate, $v_g$ is the laser group velocity, and $\mathcal F_\gamma$ is the slowly varying envelope. The center of the wakefield bubble is defined by $\zeta_0=0$. In the blowout regime~\cite{kostyukov_phenomenological_2004,lu_nonlinear_2006}, the normalized wakefield can be represented locally by
\begin{equation}
    E_{0,x}^{(w)}=\frac{k_p\zeta_0}{2}\frac{\omega_p}{\omega_0},
    \qquad
    E_{0,r}^{(w)}=-B_{0,\theta}^{(w)}
    =\frac{k_pr}{4}\frac{\omega_p}{\omega_0},
    \label{eq:wake_cylindrical}
\end{equation}
where the superscript $(w)$ denotes the wakefield, $r=(y^2+z^2)^{1/2}$ is the cylindrical radius, and $\omega_p$ and $k_p=\omega_p$ are the plasma frequency and wave number. The corresponding plasma wavelength is $\lambda_p=2\pi/k_p$. The fields are normalized to $m_e\omega_0/e$, where $m_e$ and $e$ are the electron mass and elementary charge, respectively. The equivalent Cartesian components used in the derivation are given in Appendix~\ref{app:source_derivation}.

\subsection{Axion source and coherent growth}

Combining Eqs.~(\ref{eq:laser_field}) and (\ref{eq:wake_cylindrical}) for axion generation in Eq.~(\ref{eq:axion_KG}), the dominant laser-wakefield interaction term is
\begin{equation}
    S_a\equiv\gagg\bm E_0\cdot\bm B_0
    =\frac{\gagg a_0\omega_pk_pz}{4\omega_0}
    \mathcal F_\gamma\ee^{\mi\xi_0}+\cc.
    \label{eq:axion_source}
\end{equation}
Such source is odd-symmetric with respect to $z\rightarrow-z$ and carries the laser frequency. Then the axion field can be written as
\begin{equation}
    \phi=\varphi(x)\mathcal F_a(\zeta_0,y,z)\ee^{\mi\xi_a}+\cc,
    \qquad
    \mathcal F_a=\frac{\gagg a_0\omega_pk_pz}{4\omega_0}\mathcal F_\gamma,
    \label{eq:axion_ansatz}
\end{equation}
where $\varphi(x)$ is the longitudinal amplitude, $\mathcal F_a$ is the axion envelope, and $\xi_a=k_ax-\omega_at$ is the axion phase, with axion wave number $k_a$ and frequency $\omega_a=\omega_0$. The laser and axion share a frequency but have different wave numbers. Under the slowly varying envelope approximation, $\varphi''\ll k_a\varphi'$, Eq.~(\ref{eq:axion_KG}) becomes
\begin{equation}
    -2\mi k_a\varphi'(x)=\ee^{\mi\kappa x},
    \qquad \kappa\equiv k_0-k_a.
\end{equation}
With no initially incident axion at $x=0$,
\begin{equation}
    \phi=-\frac{2\mathcal F_a}{k_a\kappa}
    \sin\left(\frac{\kappa x}{2}\right)
    \sin\left(\xi_a+\frac{\kappa x}{2}\right).
    \label{eq:axion_solution}
\end{equation}
Thus $\phi\propto x$ and the axion energy scales as $x^2$ for $|\kappa x|\ll1$. Away from phase matching, coherent growth is reduced by
\begin{equation}
    \mathcal M_a(x)=\sinc\left(\frac{\kappa x}{2}\right),
    \qquad \text{here,}\quad \sinc u\equiv\frac{\sin u}{u}.
    \label{eq:Ma}
\end{equation}

\subsection{AREM generation and frequency structure}

When the axions are generated, they interact with the laser and  the wakefields, generating secondary electromagnetic field. We calculate the AREM by the vector potential $\bm A_1$ through $\bm E_1=-\partial_t\bm A_1$ and $\bm B_1=\nabla\times\bm A_1$ from Eq.~(\ref{eq:faraday_1}-\ref{eq:ampere_1}) and one can get
\begin{equation}
    (\partial_t^2-\nabla^2)\bm A_1=\bm S_1,
    \qquad
    \bm S_1=\gagg\left[(\partial_t\phi)\bm B_0
    -\bm E_0\times\nabla\phi\right].
    \label{eq:A1_wave}
\end{equation}
The $y$-polarized source contains only the fundamental frequency, while the $z$-polarized source contains three terms. Retaining their leading transverse envelopes while suppressing common coefficients and phase-mismatch factors, they scale as
\begin{equation}
    S_{1,y}\sim z^2\mathcal F_\gamma
    \cos(\xi_a+\kappa x/2),
    \label{eq:sy_em1}
\end{equation}
\begin{equation}
    \begin{aligned}
    S_{1,z}\sim{}&yz\mathcal F_\gamma
    \cos(\xi_a+\kappa x/2),\\
    &z\mathcal F_\gamma^2
    \cos(\xi_a-\xi_0+\kappa x/2),\\
    &z\mathcal F_\gamma^2
    \cos(\xi_a+\xi_0+\kappa x/2).
    \end{aligned}
    \label{eq:sz_em1}
\end{equation}
The three terms in $S_{1,z}$ generate $z$-polarized AREM with frequency components at $\omega_0$, $0$, and $2\omega_0$, respectively. The components at $0$ and $2\omega_0$ result from the coupling between the axion field and the intense laser. Unlike conventional LSW experiments, where axion-photon conversion occurs only in static magnetic fields, our scheme combines plasma wakefields with a co-propagating intense laser pulse. In this regime the laser fields are much stronger than the effective wakefield background, so the axion reconversion is dominated not only by the wakefields but also by its coupling to the intense laser fields. This additional channel leads to AREM containing a second-harmonic component of the driving laser, a feature absent in purely static-field setups.

For any source component like
\begin{equation}
    S_1=\mathcal F_1\sin(\kappa x/2)
    \ee^{\mi(\xi^{(s)}+\kappa x/2)}+\cc,
\end{equation}
where $\mathcal F_1$ is its transverse envelope and $\xi^{(s)}=k^{(s)}x-\omega_1t$ is the source phase. The superscript $(s)$ labels a source quantity with $\xi^{(s)}$ could be $\xi_a$, $\xi_a-\xi_0$, or $\xi_a+\xi_0$ (See Eq.~(\ref{eq:sz_em1})). The corresponding vector-potential component, with phase $\xi_1=k_1x-\omega_1t$ and wave number $k_1$, is
\begin{equation}
    A_1=-\frac{2\mathcal F_1}{k_1\kappa'}
    \sin\left(\frac{\kappa x}{2}\right)
    \sin\left(\frac{\kappa'x}{2}\right)
    \sin\left(\xi_1+\frac{\kappa'x}{2}\right),
    \label{eq:A1_general}
\end{equation}
where
\begin{equation}
    \kappa'=k^{(s)}+\frac{\kappa}{2}-k_1.
    \label{eq:kappa_prime_general}
\end{equation}
Here $\kappa'$ is the mismatch between the effective source wave number $k^{(s)}+\kappa/2$ and the AREM wave number $k_1$.
Consequently, $A_1\propto\gagg^2x^2$ when both mismatch phases are small. The full component expressions are given in Appendix~\ref{app:source_derivation}.

\subsection{Transverse modes of AREM}
Since the wakefields exhibit a cylindrically symmetric structure with a longitudinal electric field and an azimuthal magnetic field, and the injected laser pulse is polarized along the $y$ direction, the generated axion field would be inverse in the $z>0$ and $z<0$ regions. Such a transverse profile would be inherited by the AREM. Therefore, besides the characteristic frequencies of the AREM, we can also predict that the AREM show special transverse profiles. To describe the transverse profile, we decompose the fields to different orders of Laguerre-Gaussian (LG) modes:
\begin{gather}
    u_{pl} = a_{pl} \frac{1}{w_\bot}\left(\frac{\sqrt{2}r_\bot}{w_\bot}\right)^{\left|l\right|} L_p^{\left|l\right|}\left(\frac{2r_\bot^2}{w_\bot^2}\right)e^{-r_\bot^2/w_\bot^2}e^{-\mi l\theta},
\end{gather}
where $r_\bot=(y^2+z^2)^{1/2}$ and $\theta$ are the transverse polar coordinates, $w_\bot$ is the LG mode width, $a_{pl}=\sqrt{\frac{2p!}{\pi(p+\left|l\right|)!}}$ is the normalization factor, $L_p^{\left|l\right|}$ is the generalized Laguerre polynomial, and $l$ and $p$ are the azimuthal and radial indices, respectively. The LG coefficient $\mathcal C_{pl}$ of a field $E$ is
$
    \mathcal{C}_{pl} = \int_0^{2\pi}\dd\theta\int_0^\infty r_\bot\dd r_\bot u_{pl}^* E(r_\bot, \theta).
$

According to the results in Eqs. (\ref{eq:sy_em1}-\ref{eq:sz_em1}), we can obtain the LG decomposition of the different components of the AREM:
\begin{alignat}{3}
    &{A}_{1, y, \omega_0}&&: \quad \mathcal{C}_{pl}\ne 0 \quad &&\text{only when} \quad l=0,\,\pm 2, \\
    &{A}_{1, z, 0\omega_0}&&: \quad \mathcal{C}_{pl}\ne 0 \quad &&\text{only when} \quad l=\pm 1, \\
    &{A}_{1, z, \omega_0}&&: \quad \mathcal{C}_{pl}\ne 0 \quad &&\text{only when} \quad l=\pm 2, \\
    &{A}_{1, z, 2\omega_0}&&: \quad \mathcal{C}_{pl}\ne 0 \quad &&\text{only when} \quad l=\pm 1.
\end{alignat}

These frequency and transverse-mode predictions provide the analytical benchmarks for the simulations below.

\section{Results from self-consistent Numerical simulations}
\label{sec:simulations}

\subsection{Simulation setup and parameters}

In the simulation, the laser has a typical wavelength $\lambda_0=800\,\mathrm{nm}$ and a normalized amplitude $a_0=5$. Its Gaussian envelope is $\mathcal F_\gamma=\exp\left[-\frac{\zeta_0^2}{w_l^2}-\frac{y^2+z^2}{w_\perp^2}\right]$, where $w_l$ and $w_\perp$ are the longitudinal and transverse envelope widths, respectively. The longitudinal duration is $12\tau_0=24\pi/\omega_0$, where $\tau_0=2\pi/\omega_0$ is the laser period, and $w_\perp=50\lambda_0$. The large spot reduces diffraction of the co-propagating axion pulse. The laser is guided in a parabolic channel, $n_e(r)=n_{e0}+\delta n\,r^2$, with $n_{e0}=0.003n_c$ and $\delta n=4\varepsilon_0m_ec^2/(e^2w_\perp^4)$, where $n_{e0}$ is the on-axis electron density, and $n_c=\varepsilon_0m_e\omega_0^2/e^2$ is the critical plasma density. As for the axions, the typical parameters we consider are $m_a=10^{-6}\,\text{eV}$ and $\gagg = 10^{-9}\,\text{GeV}^{-1}$.

To simplify the analysis, here we focus on the properties of the axion and AREM generated in the laser-wakefield interaction and neglect the plasma effect on the AREM. Therefore, the effect of different $\gagg$ is clear that the axion field $\phi\propto \gagg$ and the AREM $\bm E_1\propto \gagg^2$. Only one typical $\gagg$ is needed in a specific simulation to show the properties of the axion field and AREM. Moreover, a larger axion mass (close to the photon energy) may lead to the axion-photon phase-mismatching, and we will discuss it later.

The numerical implementation extends the code EPOCH with the axion equation and the first-order Maxwell system~\cite{arber_contemporary_2015,an_modeling_2024}. Three-dimensional Cartesian simulations resolve the detailed transverse field structure and provide the field, frequency, and LG diagnostics presented below. For centimeter propagation, simultaneously suppressing longitudinal numerical dispersion and resolving the transverse modes is expensive. We therefore also use a cylindrical implementation in which the axial and radial coordinates are discretized while the azimuthal dependence is expanded in Fourier modes~\cite{an_xdispersionless_2026}. This permits self-consistent 1-cm simulations with full plasma dynamics for both laser- and beam-driven wakefields.

\subsection{Axions and AREM produced in  Laser-driven wakefields}

\begin{figure}[t]
    \centering
    \includegraphics[width=\columnwidth]{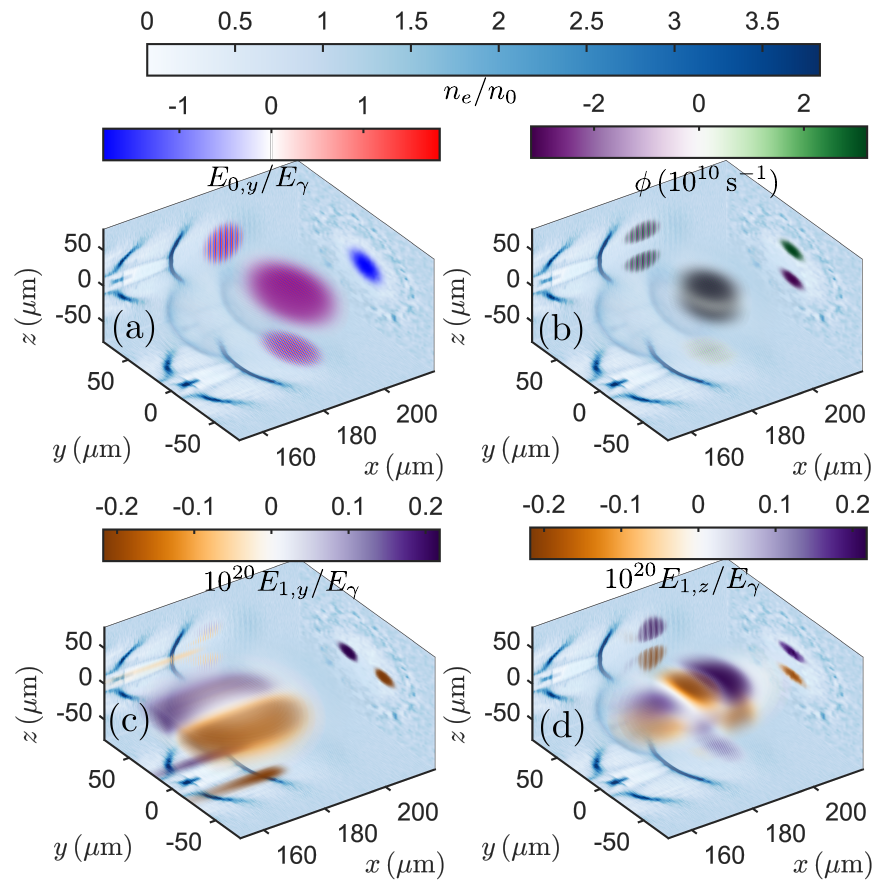}
    \caption{Typical axion and AREM fields in a laser-driven wakefield: (a) laser field $E_{0,y}$, (b) axion field $\phi$, (c) AREM component $E_{1,y}$, and (d) AREM component $E_{1,z}$. Electromagnetic fields are normalized to $E_\gamma=m_e\omega_0c/e$.}
    \label{fig:typical_fields}
\end{figure}

Figure~\ref{fig:typical_fields} shows the instantaneous, unfiltered fields from a representative Cartesian simulation. The axion and AREM fields overlap the laser pulse within the wakefield and exhibit distinct transverse symmetries.

More specifically, the node of the axion field on the $z=0$ plane and its opposite signs on the two sides of this plane in Fig.~\ref{fig:typical_fields}(b) follow directly from $S_a\propto z\mathcal F_\gamma$ in Eq.~(\ref{eq:axion_source}). The AREM appears in the same interaction region, consistent with the local source in Eq.~(\ref{eq:A1_wave}). The total $E_{1,y}$ field in Fig.~\ref{fig:typical_fields}(c) is predominantly even with respect to the $z=0$ plane, whereas the total $E_{1,z}$ field in Fig.~\ref{fig:typical_fields}(d) is predominantly odd. Because these unfiltered fields contain superposed frequency components, Fig.~\ref{fig:typical_fields} establishes their generation region and overall parity but does not isolate the individual source terms in Eqs.~(\ref{eq:sy_em1}) and (\ref{eq:sz_em1}).

\subsection{Frequency and transverse-mode signatures}

\begin{figure}[t]
    \centering
    \includegraphics[width=\columnwidth]{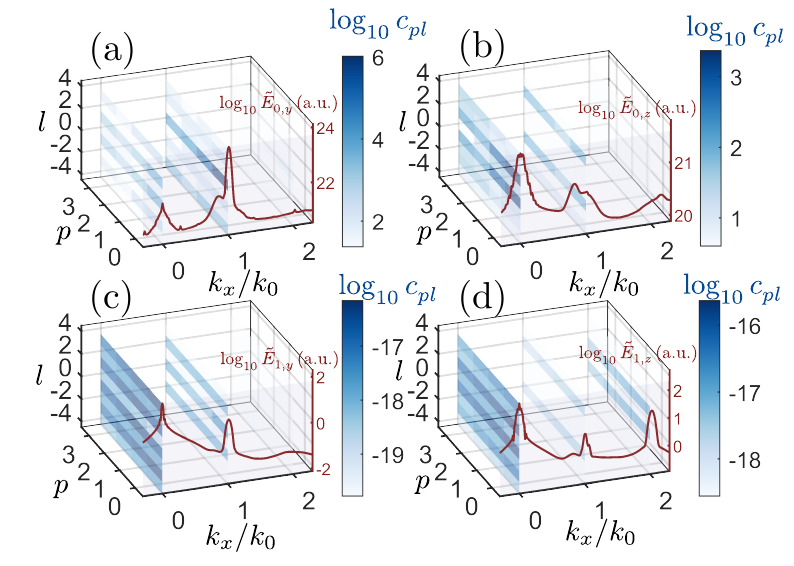}
    \caption{Frequency spectra and LG decompositions of (a) $E_{0,y}$, (b) $E_{0,z}$, (c) $E_{1,y}$, and (d) $E_{1,z}$. Red curves are frequency spectra. The mode maps show the LG coefficients of selected frequency components.}
    \label{fig:frequency_modes}
\end{figure}

The simulated spectra and LG projections of the laser and AREM fields are shown in Fig.~\ref{fig:frequency_modes}. The $E_{1,y}$ spectrum is concentrated near $\omega_0$ and contains $l=0,\pm2$, while $E_{1,z}$ exhibits the predicted zero-, fundamental-, and second-harmonic frequency components, which follow the rules of the theoretical analysis. The radial index $p$ is not restricted to a single value because the wakefield modifies the radial envelope from the chosen LG width. What remains robust is the azimuthal selection rule fixed by the symmetry.

Each simulated channel can be matched to one term in Eqs.~(\ref{eq:sy_em1}) and (\ref{eq:sz_em1}). Since $z^2=r^2\sin^2\theta$ contains the angular harmonics $l=0$ and $l=\pm2$, the peak of $E_{1,y}$ at $\omega_0$ in Fig.~\ref{fig:frequency_modes}(c) projects mainly onto these azimuthal indices. For $E_{1,z}$, the difference-frequency term $z\mathcal F_\gamma^2$ produces the near-zero-frequency peak with $l=\pm1$, the wakefield-mediated term $yz\mathcal F_\gamma\propto r^2\sin(2\theta)$ produces the fundamental peak with $l=\pm2$, and the sum-frequency term $z\mathcal F_\gamma^2$ produces the $2\omega_0$ peak with $l=\pm1$, as resolved in Fig.~\ref{fig:frequency_modes}(d). Figures~\ref{fig:frequency_modes}(a) and \ref{fig:frequency_modes}(b) show the background spectra and mode projections in these same channels.

\begin{figure}[t]
    \centering
    \includegraphics[width=\columnwidth]{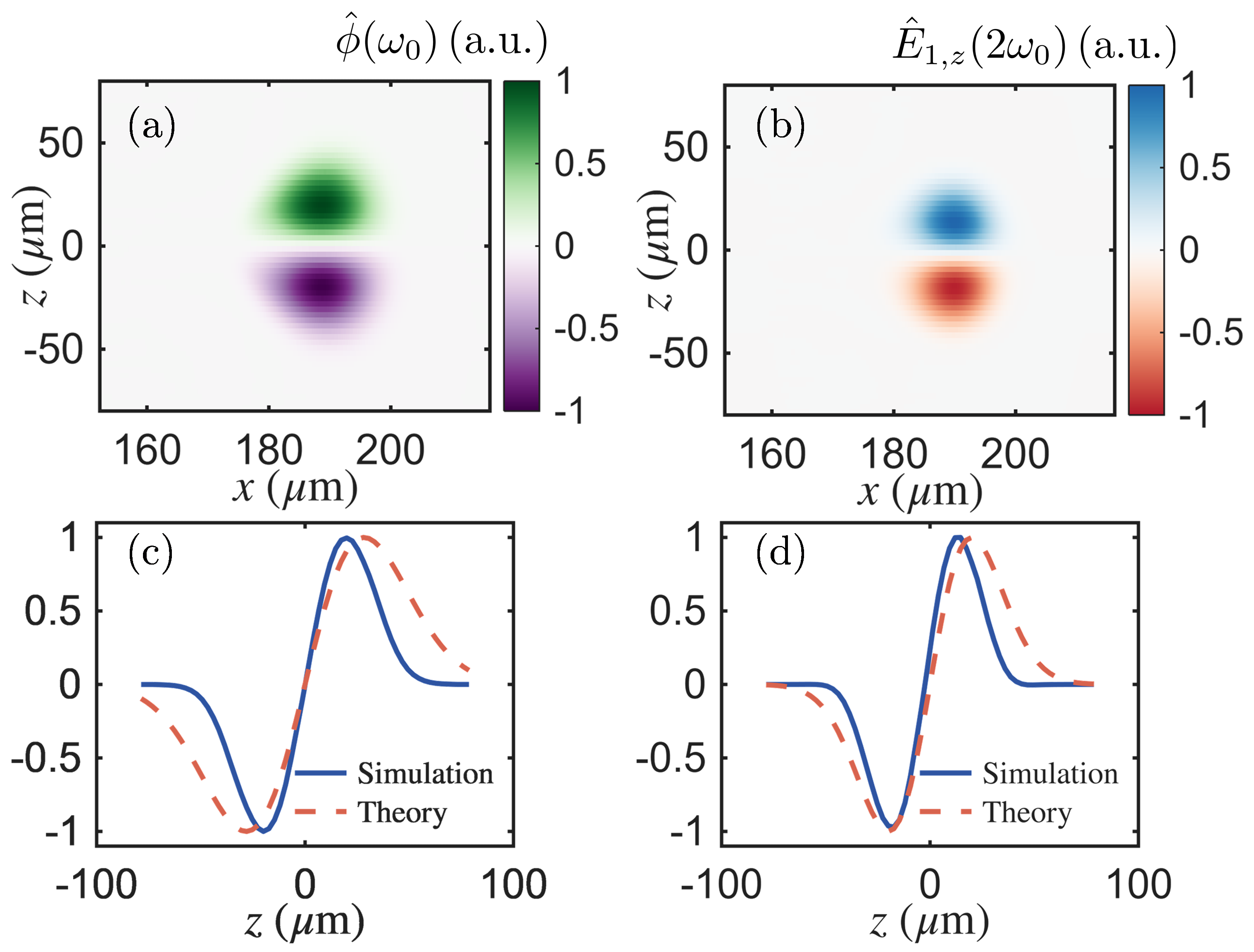}
    \caption{Frequency-resolved spatial distributions and transverse profiles in the laser-driven wakefield. (a) Axion field at $\omega_0$ on the $y=0$ plane. (b) $z$-polarized AREM at $2\omega_0$ on the $y=0$ plane. Each field in (a) and (b) is normalized to its maximum absolute value. (c) Normalized axion profile along the $z$ direction at $x=191.7\,\mu\mathrm m$. (d) Normalized $E_{1,z}(2\omega_0)$ profile at the same position. Solid curves show the simulation results. Dashed curves show the analytical envelopes proportional to $z\exp(-z^2/w_\perp^2)$ in (c) and $z\exp(-2z^2/w_\perp^2)$ in (d).}
    \label{fig:axion_E1z_profile}
\end{figure}

The frequency separation also permits a direct comparison of the simulated field envelopes with the analytical model. Figures~\ref{fig:axion_E1z_profile}(a) and \ref{fig:axion_E1z_profile}(b) show the axion field at $\omega_0$ and the $z$-polarized AREM at $2\omega_0$ on the $y=0$ plane. Both fields occupy the interaction region identified in Fig.~\ref{fig:typical_fields}. Their normalized profiles along the $z$ direction at $x=191.7\,\mu\mathrm m$ are compared with the analytical envelopes in Figs.~\ref{fig:axion_E1z_profile}(c) and \ref{fig:axion_E1z_profile}(d). The simulation reproduces the forms $z\exp(-z^2/w_\perp^2)$ for the axion and $z\exp(-2z^2/w_\perp^2)$ for $E_{1,z}(2\omega_0)$, including their parity, central node, and characteristic width. Differences in the wings arise from the self-consistent laser and wakefield profiles, which depart from the local Gaussian approximation.

\begin{figure}[t]
    \centering
    \includegraphics[width=\columnwidth]{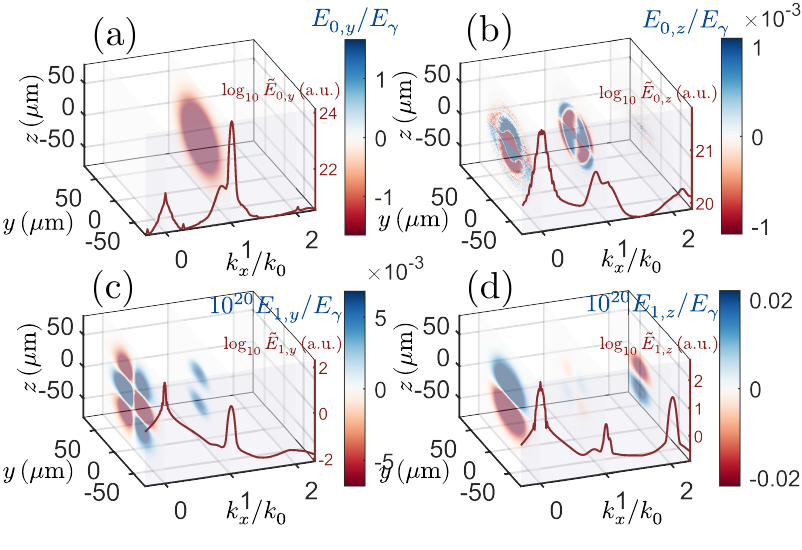}
    \caption{Transverse distributions of the frequency-resolved fields: (a) $E_{0,y}$, (b) $E_{0,z}$, (c) $E_{1,y}$, and (d) $E_{1,z}$. The petal structures provide a real-space representation of the LG selection rules.}
    \label{fig:transverse_slices}
\end{figure}

The frequency-resolved transverse slices in Fig.~\ref{fig:transverse_slices} provide a direct visualization of the azimuthal mode content identified by the LG decomposition. The $E_{1,y}(\omega_0)$ distribution in Fig.~\ref{fig:transverse_slices}(c) contains the $l=0$ and $l=\pm2$ contributions. Figure~\ref{fig:transverse_slices}(d) shows that the near-zero component of $E_{1,z}$ shares the $l=\pm1$ angular structure of the second harmonic, whereas its fundamental component forms the four-petal pattern required by $l=\pm2$. The alternating signs and nodal planes follow the transverse factors in Eq.~(\ref{eq:sz_em1}).

\section{Phase matching and driver dependence}
\label{sec:phase_matching}

\subsection{Mismatch in a partially evacuated bubble}

\begin{figure}[t]
    \centering
    \includegraphics[width=0.78\columnwidth]{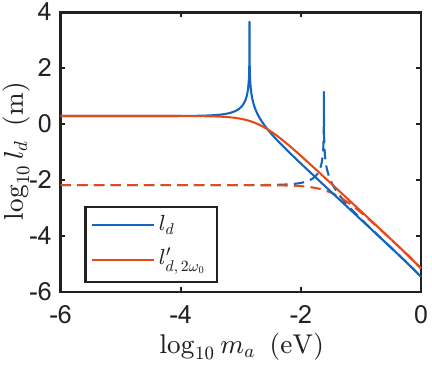}
    \caption{Axion and AREM dephasing lengths versus axion mass. Dashed curves correspond to the laser-driven bubble, whereas solid curves correspond to the cleaner beam-driven bubble. The axion-production resonance occurs near $m_a=\omega_{p,b}$.}
    \label{fig:dephasing}
\end{figure}

Long-distance accumulation requires both the axion and the AREM to remain phase matched to their sources. We define the dephasing lengths as
\begin{equation}
    l_d=\frac{2\pi}{|\kappa|},
    \qquad
    l'_{d,\mu}=\frac{2\pi}{|\kappa'_\mu|}.
    \label{eq:dephasing_lengths}
\end{equation}
where $\mu=0,1,2$ labels the AREM frequency component at $\mu\omega_0$.
The laser does not propagate in the unperturbed channel density. Electron exclusion and relativistic motion reduce the effective plasma frequency felt by the laser pulse to
\begin{equation}
    \omega_{p,b}^2=\frac{n_{\mathrm{eff}}e^2}{\gamma_\perp m_e},
    \qquad
    \gamma_\perp=\sqrt{1+a_0^2/2}.
    \label{eq:bubble_frequency}
\end{equation}
Here $\omega_{p,b}$ is the effective plasma frequency inside the bubble, $n_{\mathrm{eff}}$ is the effective residual electron density, and $\gamma_\perp$ is the relativistic transverse factor.
For the laser-driven example, the simulations give $n_{\mathrm{eff}}\simeq0.3n_{e0}$ and $\gamma_\perp\simeq3.7$. The axion-production mismatch is
\begin{equation}
    \kappa=\sqrt{\omega_0^2-\omega_{p,b}^2}
    -\sqrt{\omega_0^2-m_a^2}.
    \label{eq:kappa_dispersion}
\end{equation}
For the $2\omega_0$ AREM component,
\begin{equation}
    \kappa'_2=\frac{1}{2}\sqrt{\omega_0^2-m_a^2}
    +\frac32\sqrt{\omega_0^2-\omega_{p,b}^2}
    -\sqrt{4\omega_0^2-\omega_{p,b}^2}.
    \label{eq:kappa_prime_dispersion}
\end{equation}
Equation~(\ref{eq:kappa_dispersion}) vanishes at $m_a=\omega_{p,b}$, producing the resonance in $l_d$ shown in Fig.~\ref{fig:dephasing}, where the dephasing lengths for axion and AREM generation are shown. The AREM mismatch does not vanish at the same point because its source and emitted photon involve different frequency combinations. In the light-mass limit,
\begin{equation}
    |\kappa|\simeq|\kappa'_2|
    \simeq\frac{\omega_{p,b}^2}{2\omega_0},
    \label{eq:light_mass_mismatch}
\end{equation}
so residual electrons in the wakefield bubble would affect the dephasing lengths.

\subsection{Axion and AREM generation in a beam-driven bubble}

\begin{figure}[t]
    \centering
    \includegraphics[width=\columnwidth]{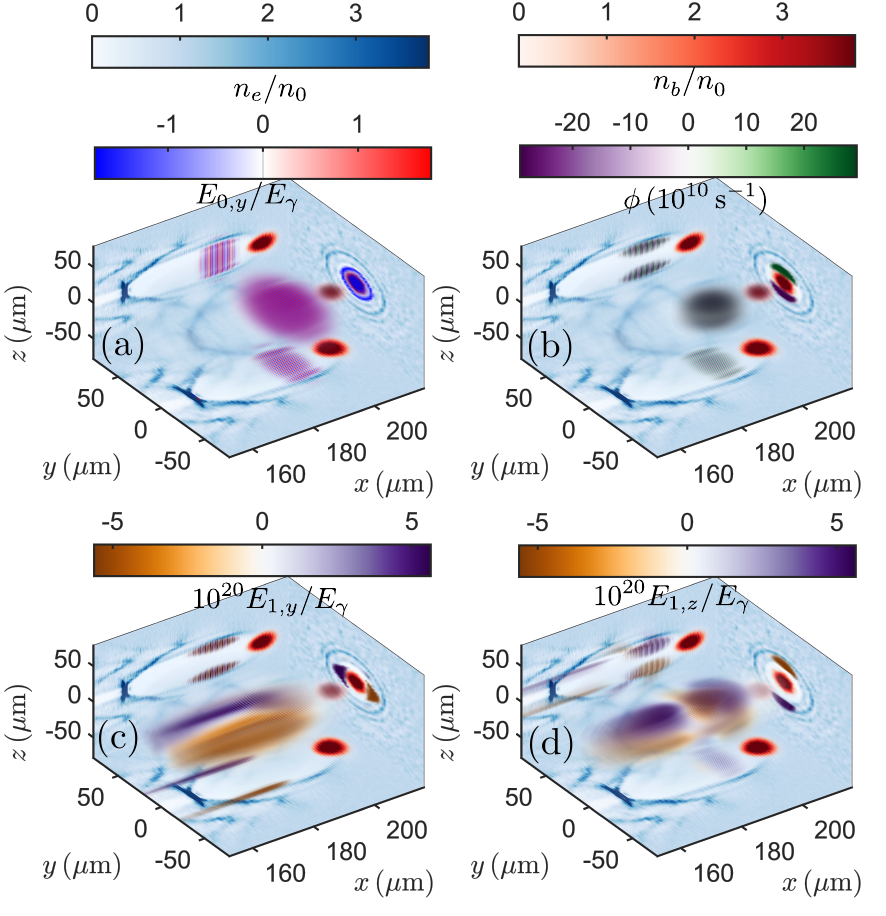}
    \caption{Axion generation and conversion in the beam-driven wakefield: (a) laser field $E_{0,y}$, (b) axion field $\phi$, (c) AREM component $E_{1,y}$, and (d) AREM component $E_{1,z}$. The electromagnetic fields are normalized to $E_\gamma$.}
    \label{fig:scheme_edrive}
\end{figure}

We compare laser- and beam-driven wakefields in simulations to examine electron evacuation inside the bubble and the associated axion and AREM generation. The electron beam to drive the wakefield has an energy of $4\,\mathrm{GeV}$, a Gaussian length of $4\,\mu\mathrm m$, a radius of $12\,\mu\mathrm m$, and a peak density of $5n_{e0}$, while the laser and plasma parameters are the same as in the laser-driven case. The laser pulse trails the electron beam by $15\,\mu\mathrm m$, rather than the linear-wakefield estimate $\lambda_p/2=7\,\mu\mathrm m$, because the wakefield is strongly nonlinear. Although the wakefield drivers are different, they produce similar wakefield structures. Consequently, the axion and AREM fields generated in the beam-driven wakefield have spatial structures similar to those in the laser-driven wakefield, as shown in Fig.~\ref{fig:scheme_edrive}.

\begin{figure}[t]
    \centering
    \includegraphics[width=\columnwidth]{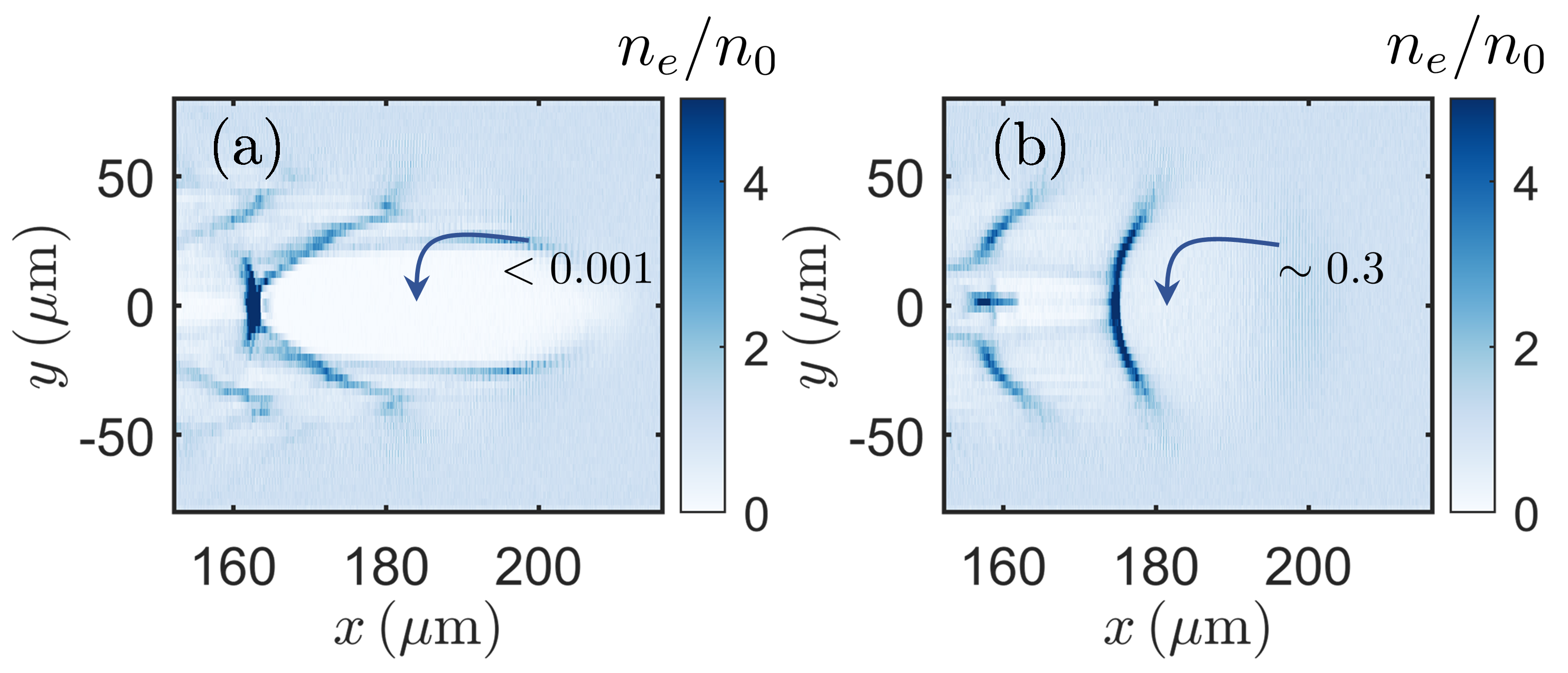}
    \caption{Electron density inside the wakefield bubble driven by (a) an electron beam and (b) a laser pulse. The lower residual density of the beam-driven bubble reduces the effective plasma frequency and extends the dephasing length.}
    \label{fig:density_bubble}
\end{figure}

As shown in Fig.~\ref{fig:density_bubble}, the residual density in the beam-driven bubble can be below $10^{-3}n_{e0}$, corresponding to $\omega_{p,b}<1.4\,\mathrm{meV}$ under the intensity-weighted dispersion estimate. Equation~(\ref{eq:light_mass_mismatch}) then gives $l_d\simeq l'_d\gtrsim2\,\mathrm m$ for axions with masses below $1\,\mathrm{meV}$, provided the same wakefield structure is maintained.

\subsection{Simulation-calibrated scaling with propagation distance}

The cleaner beam-driven bubble discussed above improves phase matching and thereby permits coherent accumulation over longer distances. We now quantify this accumulation using the theoretical scaling and the self-consistent cylindrical simulations. When phase mismatch is negligible, both the axion conversion probability and the AREM amplitude scale quadratically with the interaction length. The axion conversion probability is defined as the ratio of the axion energy to the initial laser energy,
\begin{equation}
    P_a=\frac{\int\mathcal H_a\dd V}{\int\mathcal H_\gamma\dd V},
    \label{eq:Pa_definition}
\end{equation}
where
\begin{align}
    \mathcal H_a&=\frac12\left[(\partial_t\phi)^2+(\nabla\phi)^2+m_a^2\phi^2\right],\\
    \mathcal H_\gamma&=\frac12(\bm E_0^2+\bm B_0^2).
\end{align}
$\mathcal H_a$ and $\mathcal H_\gamma$ are the axion and electromagnetic energy densities, respectively, and the integrals extend over the simulation volume $V$.
Axion production depends not only on the interaction length but also on the effective field strength experienced by the laser. The typical wakefield strength is set by the plasma frequency and hence the plasma density, with $B_p=m_e\omega_p/e$. The effective background field experienced by the laser can be stronger because of wakefield nonlinearity, the transverse width, and the contributions of both electric and magnetic fields to axion generation. These effects, particularly the wakefield nonlinearity, depend strongly on the wakefield driver and plasma parameters and do not admit a general description. We therefore introduce a phenomenological nonlinearity factor $\chi\in(1,10)$ to account for them. Including the phase-mismatch factor defined in Eq.~(\ref{eq:Ma}), the simulation-calibrated production law is
\begin{equation}
    P_a=P_s\chi^2
    \left(\frac{\gagg}{10^{-9}\,\mathrm{GeV}^{-1}}\right)^2
    \left(\frac{n_{e0}}{10^{18}\,\mathrm{cm}^{-3}}\right)
    \left(\frac{L}{1\,\mathrm{cm}}\right)^2
    \mathcal M_a^2,
    \label{eq:Pa_scaling}
\end{equation}
where $L$ is the interaction length and $P_s=1.0\times10^{-17}$ is the reference conversion ratio obtained from the simulation with $\gagg=10^{-9}\,\mathrm{GeV}^{-1}$ and $L=1\,\mathrm{cm}$. The beam-driven simulation achieves $\chi=4$. The corresponding conversion probability is shown in Fig.~\ref{fig:generation_filter}(a), where the stronger wakefield gives a larger conversion probability than the ALPS-II production region.

\section{Direct detection of the AREM}
\label{sec:direct_detection}

\begin{figure}[t]
    \centering
    \includegraphics[width=\columnwidth]{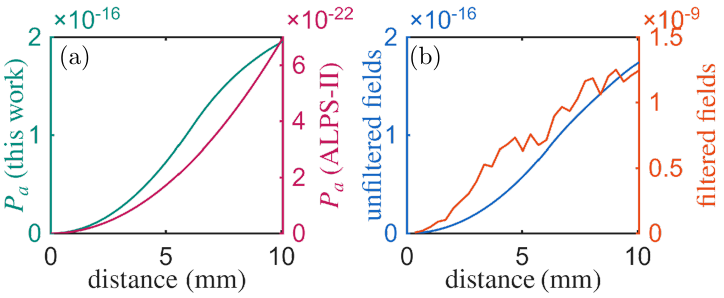}
    \caption{Simulation-calibrated production and filtering. (a) Axion conversion probability in the wakefield and in the ALPS-II production magnet. (b) AREM-to-background field ratio before filtering and after selecting $z$ polarization, $2\omega_0$, and the $|l|=1$ subspace. The results are calculated using an electron beam as wakefield driver with beam energy 40\,GeV, duration 2\,$\mu$m, radius 7\,$\mu$m, and peak density $6\times10^{20}\,\mathrm{cm}^{-3}$ to drive a clearer and stronger wakefield bubble. The laser inside the wakefield has the same parameters as before.}
    \label{fig:generation_filter}
\end{figure}

Beyond improving LSW experiments, an alternative approach is to directly detect the AREM within the interaction region. Recent experiments have demonstrated high-power laser guiding in plasma channels over meter-scale distances~\cite{miao_optical_2020,picksley_matched_2024,miao_benchmarking_2024,shrock_guided_2024}. At such extended distances, the AREM signal could become sufficiently strong for direct detection. The distinct properties of the AREM enable multi-stage filtering to significantly improve the signal-to-noise ratio (SNR), defined as
\begin{equation}
    \mathrm{SNR}_{\arem}=\frac{\overline E_1}{\overline E_0},
    \qquad
    \overline E_j=\left(\frac1V\int_V E_j^2\dd V\right)^{1/2},
    \label{eq:arem_field_ratio}
\end{equation}
where $j=0,1$ labels the background and AREM fields, respectively, $V$ is the sampled volume, and the same polarization, frequency interval, and spatial mode are applied to both fields. As shown in Fig.~\ref{fig:generation_filter}(b), one can enhance the SNR by several orders of magnitude by simultaneously selecting the $z$ polarization, $2\omega_0$ frequency, and $|l|=1$ LG modes~\cite{fontaine_laguerre-gaussian_2019}.

The filtered AREM-to-background root-mean-square (RMS) field ratio follows
\begin{equation}
    \begin{aligned}
    \frac{\overline E_1}{\overline E_0}={}&\mathcal R_s\chi^2
    \left(\frac{\gagg}{10^{-9}\,\mathrm{GeV}^{-1}}\right)^2\\
    &\times\left(\frac{n_{e0}}{10^{18}\,\mathrm{cm}^{-3}}\right)
    \left(\frac{L}{1\,\mathrm{cm}}\right)^2
    \mathcal M_a\mathcal M_2,
    \end{aligned}
    \label{eq:E1_scaling}
\end{equation}
where
\begin{equation}
    \mathcal M_a=\sinc\left(\frac{\kappa L}{2}\right),
    \qquad
    \mathcal M_2=\sinc\left(\frac{\kappa'_2L}{2}\right).
\end{equation}
Here $\mathcal R_s=1.5\times10^{-9}$ is the reference RMS field ratio obtained from the simulation with $\chi=4$, $\gagg=10^{-9}\,\mathrm{GeV}^{-1}$, and $L=1\,\mathrm{cm}$.

When phase mismatch is negligible, Eq.~(\ref{eq:E1_scaling}) shows that the SNR increases quadratically with the interaction length. If a mode-resolved RMS field ratio $\overline E_1/\overline E_0=\epsilon$ can be measured, the corresponding axion-photon coupling sensitivity is
\begin{equation}
    \begin{aligned}
    \gagg^{\mathrm{sens}}={}&10^{-9}\,\mathrm{GeV}^{-1}
    \Bigg[
    \frac{\epsilon}{\mathcal R_s\chi^2}
    \frac{10^{18}\,\mathrm{cm}^{-3}}{n_{e0}}\\
    &\times\left(\frac{1\,\mathrm{cm}}{L}\right)^2
    \frac{1}{|\mathcal M_a\mathcal M_2|}
    \Bigg]^{1/2}.
    \end{aligned}
    \label{eq:arem_sensitivity}
\end{equation}
For a generation length of $1.5\,\mathrm m$ and a detection sensitivity threshold $\epsilon=10^{-4}$, the projected axion-photon coupling constraint is shown by the AREM curve in Fig.~\ref{fig:constraints}. This sensitivity is comparable to ALPS-I and exceeds those of other current ground-based laboratory experiments.

\section{wakefield LSW detection using a seeding pulse}
\label{sec:seeded_detection}

\subsection{Coherent regeneration behind the wall}

In addition to direct AREM detection, the generated axion pulse can be detected in a wakefield light-shining-through-a-wall (WLSW) configuration, which has a much lower noise for axion detection. After passing through the wall, the axions enter a downstream magnetic region and reconvert into photons. Let $n_L$ be the number of photons in the drive pulse and $P_a$ the wakefield production probability. A magnetic region with field $B_2$ and length $l_2$ reconverts an axion with probability
\begin{equation}
    P'_a=\left(\frac{\gagg B_2l_2}{2}\right)^2\mathcal M_R^2,
    \label{eq:reconversion_probability}
\end{equation}
where $\mathcal M_R$ is its phase-mismatch factor. Without a seed photon in the regeneration region, the regenerated photon expectation per shot is
\begin{equation}
    n_{\gamma0}=n_LP_aP'_a.
    \label{eq:ngamma0}
\end{equation}
Because $P_a\propto\gagg^2$ and $P'_a\propto\gagg^2$, $n_{\gamma0}\propto\gagg^4$.

To increase the regenerated photon number, following Ref.~\cite{an_coherently_2026}, a phase-controlled seed pulse is injected into the regeneration region synchronously with the axion pulse. The seed coherently interferes with the axion-induced electromagnetic field, giving the output photon number as
\begin{equation}
    N'_\gamma=n_s+n_{\gamma0}
    +2\sqrt{n_sn_{\gamma0}}\cos\delta,
    \label{eq:seed_output}
\end{equation}
where $n_s$ is the mean seed photon number and $\delta$ is the relative phase between the seed and regenerated fields. The net change attributed to the axion is
\begin{equation}
    \Delta N=n_{\gamma0}
    +2\sqrt{n_sn_{\gamma0}}\cos\delta.
    \label{eq:seed_deltaN}
\end{equation}
The seed acts as a local reference field that converts a small field amplitude into a larger intensity difference.

Writing $\mathcal C_\delta\equiv\langle\cos\delta\rangle$ and assuming $n_s\gg n_{\gamma0}$, let $R$ be the shot repetition rate and $T_{\mathrm{run}}$ the total running time. We require one expected observed signal photon over the full run,
\begin{equation}
    RT_{\mathrm{run}}\left(n_{\gamma0}
    +2\mathcal C_\delta\sqrt{n_sn_{\gamma0}}\right)=1.
    \label{eq:one_photon_criterion}
\end{equation}
In the seed-dominated limit this gives
\begin{equation}
    \begin{aligned}
    \gagg^{\mathrm{sens}}\simeq{}&
    \sqrt{2}\left(R^2T_{\mathrm{run}}^2n_Ln_s\right)^{-1/4}\\
    &\times\left(\chi B_pl_1B_2l_2
    \mathcal M_a\mathcal M_R\mathcal C_\delta\right)^{-1/2},
    \end{aligned}
    \label{eq:ideal_seeded_sensitivity}
\end{equation}
where $l_1$ is the wakefield production length; $B_2$ and $l_2$ are the regeneration magnetic field and length defined above. The coherence $\mathcal C_\delta$ is determined by the joint seed-number and phase uncertainties considered below.

\subsection{Fluctuation effects of seed photon number and phase}
\label{sec:SG_model}

We model the seed photon number and relative phase as independent Gaussian random variables,
\begin{equation}
    n\sim\mathcal N(n_s,\sigma_n^2),
    \qquad
    \delta\sim\mathcal N(0,\sigma_\delta^2),
    \qquad
    \sigma_n\ll n_s.
    \label{eq:Gaussian_seed}
\end{equation}
Here $n$ is the photon number in an individual seed pulse, while $\sigma_n$ and $\sigma_\delta$ are the standard deviations of the seed photon number and relative phase, respectively.
The interference observable per shot is
\begin{equation}
    X=2\sqrt{n n_{\gamma0}}\cos\delta.
    \label{eq:X_definition}
\end{equation}
We can evaluate the expectation value of $X$ by integrating over the joint probability distribution of $n$ and $\delta$. The result is,
\begin{equation}
    \langle X\rangle\simeq
    2\sqrt{n_{\gamma0}n_s}\ee^{-\sigma_\delta^2/2}.
    \label{eq:X_average}
\end{equation}
For $N_{\mathrm{shot}}=RT_{\mathrm{run}}$ statistically independent shots, the signal-to-noise ratio is defined as
\begin{equation}
    \mathcal S=\sqrt{N_{\mathrm{shot}}}
    \frac{2\sqrt{n_{\gamma0}n_s}\ee^{-\sigma_\delta^2/2}}
    {\sqrt{\sigma_n^2+n_b}}.
    \label{eq:working_snr}
\end{equation}
Here $n_b$ is the mean number of background photons per shot, statistically independent of the seed preparation. Assuming Poisson background counts, their per-shot variance is $n_b$. The seed-number variance $\sigma_n^2$ and background variance $n_b$ therefore add in Eq.~(\ref{eq:working_snr}).

Optical phase does not possess a unique Hermitian operator canonically conjugate to photon number. We use the Susskind--Glogower (SG) exponential-phase construction~\cite{susskind_quantum_1964,carruthers_phase_1968}, which yields
\begin{equation}
    \Delta n^2\,\Delta(\sin\delta)^2
    \geq\frac14|\langle\cos\delta\rangle|^2.
    \label{eq:SG_relation}
\end{equation}
For the Gaussian phase distribution,
\begin{equation}
    \Delta(\sin\delta)^2=\frac{1-\ee^{-2\sigma_\delta^2}}{2},
    \qquad
    |\langle\cos\delta\rangle|^2=\ee^{-\sigma_\delta^2}.
    \label{eq:Gaussian_phase_moments}
\end{equation}
When Eq.~(\ref{eq:SG_relation}) is saturated, defining $y=\ee^{-\sigma_\delta^2}$ gives the physical root
\begin{equation}
    y(\sigma_n)=
    \frac{\sqrt{1+16\sigma_n^4}-1}{4\sigma_n^2}
    =\frac{4\sigma_n^2}{\sqrt{1+16\sigma_n^4}+1}.
    \label{eq:y_SG}
\end{equation}
The corresponding SNR is
\begin{equation}
    \mathcal S_{\mathrm{SG}}(\sigma_n)=
    \sqrt{N_{\mathrm{shot}}}
    \frac{2\sqrt{n_{\gamma0}n_s}\,y(\sigma_n)^{1/2}}
    {\sqrt{\sigma_n^2+n_b}}.
    \label{eq:S_SG}
\end{equation}

\begin{figure}[t]
    \centering
    \includegraphics[width=\columnwidth]{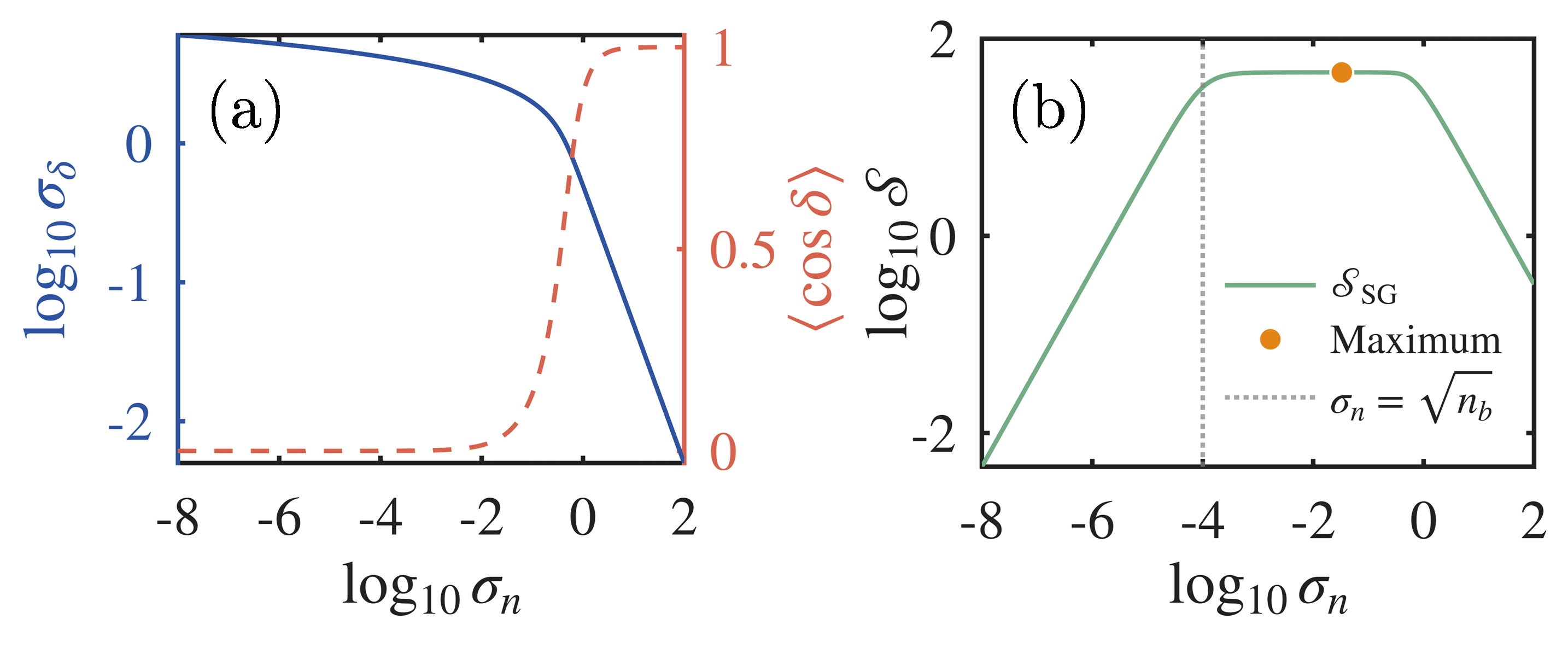}
    \caption{Joint seed-number and phase uncertainty under the saturated SG relation. (a) Gaussian phase width $\sigma_\delta$ and coherence $\langle\cos\delta\rangle=\exp(-\sigma_\delta^2/2)$ versus number uncertainty. (b) SNR from Eq.~(\ref{eq:S_SG}) for $n_{\gamma0}=10^{-6}$, $n_s=100$, $N_{\mathrm{shot}}=30\,\mathrm{days}\times1\,\mathrm{Hz}$, and a mean background of $n_b=10^{-8}$ photons per shot. The marker denotes the numerical maximum. The dotted line marks equal seed and background variances, $\sigma_n=\sqrt{n_b}$.}
    \label{fig:SG_tradeoff}
\end{figure}

Figure~\ref{fig:SG_tradeoff} shows the resulting number-phase tradeoff and SNR. As $\sigma_n\rightarrow0$,
\begin{equation}
    y\simeq2\sigma_n^2,
    \qquad
    \sigma_\delta^2\simeq-\ln(2\sigma_n^2),
    \label{eq:SG_small_sigma}
\end{equation}
so perfect number stabilization destroys the phase reference. If $n_b=0$, the loss of coherence and the reduction in number noise cancel, leaving
\begin{equation}
    \mathcal S_{\mathrm{SG}}\rightarrow
    2\sqrt{2N_{\mathrm{shot}}n_{\gamma0}n_s}.
\end{equation}
For any nonzero $n_b$ under the convention of Eq.~(\ref{eq:working_snr}), however,
\begin{equation}
    \mathcal S_{\mathrm{SG}}\simeq
    \frac{2\sqrt{2N_{\mathrm{shot}}n_{\gamma0}n_s}}{\sqrt{n_b}}\sigma_n
    \rightarrow0.
    \label{eq:S_small_sigma_nb}
\end{equation}
For $1\ll\sigma_n\ll n_s$, the coherence approaches unity but the denominator grows as $\sigma_n$, so the SNR again decreases. A finite optimum follows when $n_b>0$ within the adopted domain.

Equation~(\ref{eq:S_SG}) factorizes as
\begin{equation}
    \mathcal S_{\mathrm{SG}}(\sigma_n;n_s)
    =\sqrt{n_s}\,K\,F_{\mathrm{SG}}(\sigma_n;n_b),
    \label{eq:S_factorization}
\end{equation}
where
\begin{equation}
    K=2\sqrt{N_{\mathrm{shot}}n_{\gamma0}},
    \qquad
    F_{\mathrm{SG}}(\sigma_n;n_b)
    =\sqrt{\frac{y(\sigma_n)}{\sigma_n^2+n_b}}.
    \label{eq:SG_factors}
\end{equation}
Thus, $K$ is independent of $\sigma_n$ and $n_s$. For fixed $n_b$, increasing $n_s$ shifts the SNR curve vertically but does not move its maximum or change its logarithmic shape.

\begin{figure}[!t]
    \centering
    \includegraphics[width=0.78\columnwidth]{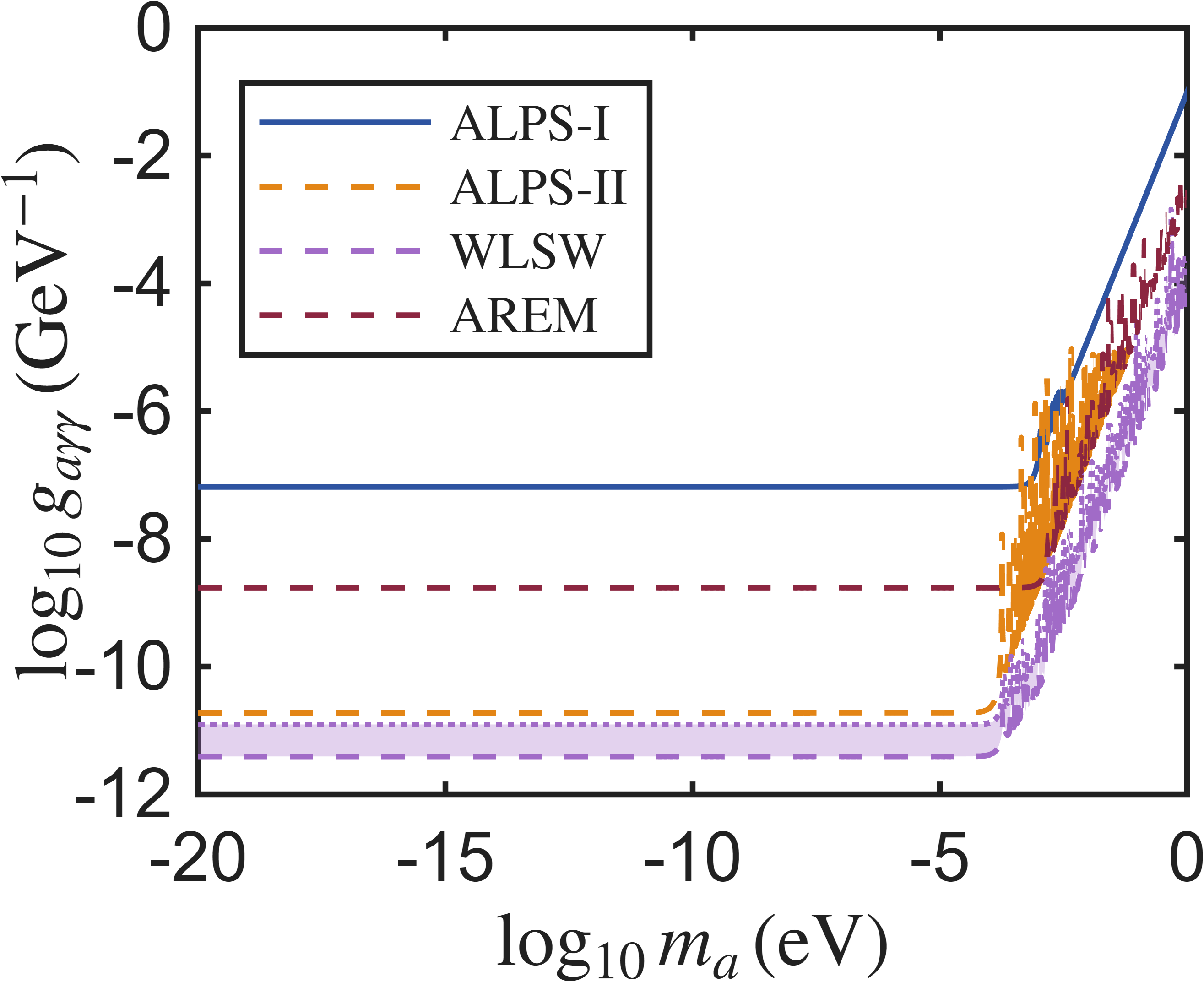}
    \caption{The axion-photon coupling constraint that can be achieved in this work. The solid lines show the constraints from the ALPS-I experiments~\cite{AxionLimits}, while the dashed lines show the constraints from this work and ALPS-II. The purple shaded region indicates the detection region for $\chi$ within $(1,\, 10)$. }
    \label{fig:constraints}
\end{figure}

For $n_b=10^{-8}$, maximizing Eq.~(\ref{eq:S_SG}) gives
\begin{equation}
    \sigma_{n,\mathrm{opt}}=3.28\times10^{-2},\qquad
    \sigma_{\delta,\mathrm{opt}}=2.48,
\end{equation}
and hence $\mathcal C_{\delta,\mathrm{opt}}=4.64\times10^{-2}$. We use this optimized coherence in Eq.~(\ref{eq:one_photon_criterion}) to evaluate the constraint on $\gagg$. As shown in Fig.~\ref{fig:constraints}, for $R=1\,\mathrm{Hz}$, $T_{\mathrm{run}}=30$ days, $n_L=4\times10^{20}$, $n_{e0}=5\times10^{18}\,\mathrm{cm}^{-3}$, $l_1=1.5\,\mathrm m$, $B_2l_2=500\,\mathrm{T\,m}$, and $n_s=100$, the low-mass WLSW band reaches $3.92\times10^{-12}$--$1.24\times10^{-11}\,\mathrm{GeV}^{-1}$ for $\chi\in(1,\,10)$.
This WLSW band lies below the ALPS-II projection in the low-mass region. The same figure compares the WLSW and AREM sensitivities derived in Secs.~\ref{sec:seeded_detection} and \ref{sec:direct_detection} with existing laboratory LSW limits~\cite{AxionLimits}.

\section{Conclusion}
\label{sec:conclusion}

In this work, we have proposed a promising platform for in situ axion generation and conversion in plasma wakefields.
The strong laser fields in plasma wakefields can drive axion production via the Primakoff effect.
For the first time, self-consistent PIC simulations have been carried out to illustrate the axion production and the re-conversion process in the wakefields.
To overcome the limitations of cavity-free short-pulse detection, we have integrated a seeded photon scheme into the LSW setup.
Assuming 500 $\text{T}\cdot\text{m}$ at the detection stage with a 100-photon seed, one can achieve a highly competitive level of $\gagg \sim 10^{-12}\,\text{GeV}^{-1}$ for axion mass less than 0.1\,meV, surpassing the bounds of next-generation cavity experiments.
We have also investigated the features of the AREM, which show unique properties on the polarization, frequency, and transverse distribution.
These features can be used to filter the AREM from the background fields, enabling a complementary promising way to detect the axions produced in the wakefields.

\appendix

\section{The source terms for axion and AREM generation}
\label{app:source_derivation}

In this appendix, we provide the explicit source terms for AREM generation and derive the corresponding vector potentials. These expressions complement the frequency and transverse-mode analysis in Sec.~\ref{sec:theory} by retaining the coefficients and phase-mismatch factors associated with both stages of the process: axion generation and subsequent axion-to-photon conversion. The Cartesian form of Eq.~(\ref{eq:wake_cylindrical}) is
\begin{align}
    E_{0,x}^{(w)}&=\frac{k_p\zeta_0}{2}\frac{\omega_p}{\omega_0},\\
    E_{0,y}^{(w)}=-B_{0,z}^{(w)}
    &=\frac{k_py}{4}\frac{\omega_p}{\omega_0},\\
    E_{0,z}^{(w)}=B_{0,y}^{(w)}
    &=\frac{k_pz}{4}\frac{\omega_p}{\omega_0},
\end{align}
with $B_{0,x}^{(w)}=0$. Inserting this wakefield and Eq.~(\ref{eq:axion_solution}) into Eq.~(\ref{eq:A1_wave}) gives
\begin{equation}
    S_{1,x}=0,
\end{equation}
\begin{equation}
    S_{1,y}=
    \frac{\gagg\mathcal F_a(k_a+\omega_a)\omega_pk_pz}
    {2k_a\kappa\omega_0}
    \sin\left(\frac{\kappa x}{2}\right)
    \cos\left(\xi_a+\frac{\kappa x}{2}\right),
    \label{eq:app_S1y}
\end{equation}
and
\begin{equation}
\begin{aligned}
    S_{1,z}={}&-
    \frac{\gagg\mathcal F_a}{2k_a\kappa\omega_0}
    \sin\left(\frac{\kappa x}{2}\right)
    \Bigg\{
    (k_a+\omega_a)\omega_pk_py\\
    &\times\cos\left(\xi_a+\frac{\kappa x}{2}\right)
    -2a_0\mathcal F_\gamma\kappa\omega_0\\
    &\times\Bigg[
    \cos\left(\xi_a-\xi_0+\frac{\kappa x}{2}\right)\\
    &\quad+
    \cos\left(\xi_a+\xi_0+\frac{\kappa x}{2}\right)
    \Bigg]\Bigg\}.
\end{aligned}
\label{eq:app_S1z}
\end{equation}

Let $\mu=0,1,2$ label the emitted frequencies $\omega_\mu^{(1)}=\mu\omega_0$ and source wave numbers
\begin{equation}
    k_\mu^{(s)}=k_a-k_0,\quad k_a,\quad k_a+k_0,
\end{equation}
respectively. With $\xi_\mu^{(1)}=k_\mu^{(1)}x-\omega_\mu^{(1)}t$ and
\begin{equation}
    \kappa'_\mu=k_\mu^{(s)}-k_\mu^{(1)}+\frac{\kappa}{2},
\end{equation}
the vector potentials are
\begin{equation}
\begin{split}
    A_{1,y}={}&-
    \frac{\gagg\mathcal F_a(k_a+\omega_a)\omega_pk_pz}
    {k_a\kappa k_1^{(1)}\kappa'_1\omega_0}
    \sin\left(\frac{\kappa x}{2}\right)
    \sin\left(\frac{\kappa'_1x}{2}\right)\\
    &\times\sin\left(\xi_1^{(1)}+\frac{\kappa'_1x}{2}\right),
\end{split}
\label{eq:app_A1y}
\end{equation}
and
\begin{equation}
\begin{aligned}
    A_{1,z}={}&-
    \frac{\gagg\mathcal F_a}{k_a\kappa\omega_0}
    \sin\left(\frac{\kappa x}{2}\right)
    \Bigg\{
    \frac{(k_a+\omega_a)\omega_pk_py}
    {k_1^{(1)}\kappa'_1}\\
    &\times\sin\left(\frac{\kappa'_1x}{2}\right)
    \sin\left(\xi_1^{(1)}+\frac{\kappa'_1x}{2}\right)\\
    &-2a_0\mathcal F_\gamma\kappa\omega_0
    \Bigg[
    \frac{\sin(\kappa'_0x/2)}{k_0^{(1)}\kappa'_0}\\
    &\quad\times
    \sin\left(\xi_0^{(1)}+\frac{\kappa'_0x}{2}\right)
    +\frac{\sin(\kappa'_2x/2)}{k_2^{(1)}\kappa'_2}\\
    &\quad\times
    \sin\left(\xi_2^{(1)}+\frac{\kappa'_2x}{2}\right)
    \Bigg]\Bigg\}.
\end{aligned}
\label{eq:app_A1z}
\end{equation}
These expressions make the two-stage coherent growth and the $\gagg^2$ scaling explicit.

\begin{figure*}[t]
    \begin{minipage}[t]{0.48\textwidth}
        \centering
        \includegraphics[width=\columnwidth]{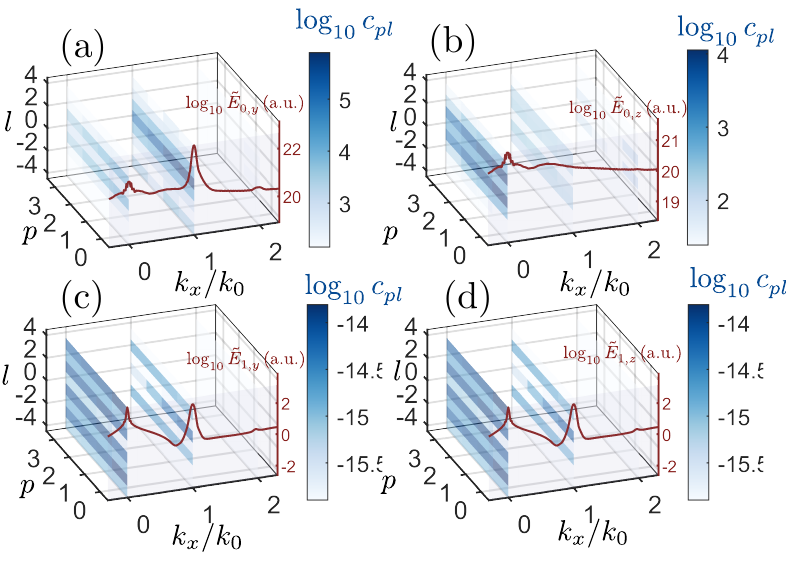}
        \caption{Frequency spectra and LG decompositions of the background and AREM in the beam-driven wakefield. The frequency components and azimuthal selection rules agree with the laser-driven case.}
        \label{fig:beam_mode_appendix}
    \end{minipage}
    \hfill
    \begin{minipage}[t]{0.48\textwidth}
        \centering
        \includegraphics[width=\columnwidth]{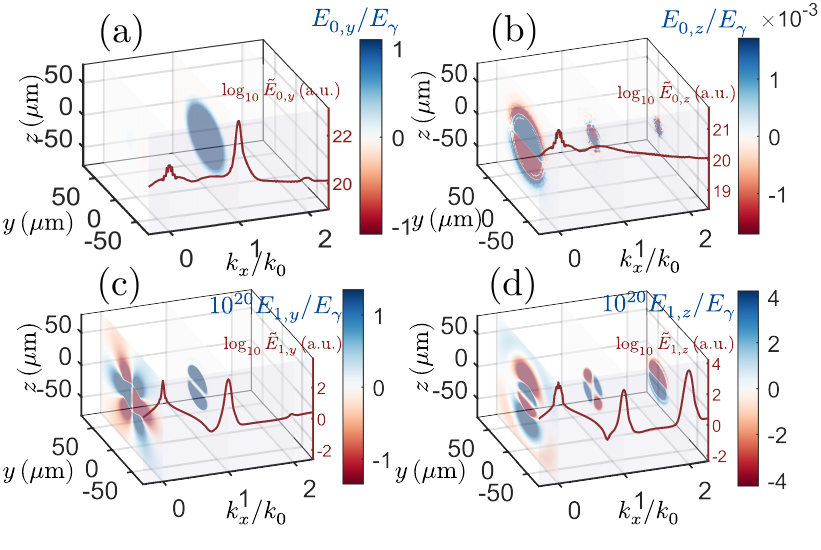}
        \caption{Transverse distributions of the frequency-resolved background and AREM components in the beam-driven wakefield. The field structures are similar to those in the laser-driven case.}
        \label{fig:beam_slice_appendix}
    \end{minipage}
\end{figure*}

The beam-driven wakefield preserves the same blowout-field geometry and laser polarization as the laser-driven case. Consequently, the frequency spectra and LG decompositions in Fig.~\ref{fig:beam_mode_appendix} exhibit the same characteristic channels and azimuthal selection rules. The frequency-resolved slices in Fig.~\ref{fig:beam_slice_appendix} also retain the corresponding parity and transverse structures. Changing the wakefield driver therefore modifies the field amplitudes and phase matching but not these qualitative signatures of the axion and AREM fields.

\section*{Acknowledgments}

This work was supported by the National Natural Science Foundation of China (Grants No. 12225505, 12090060, and 12090061), Office of Science and Technology and Shanghai Municipal Government (Grant No. 23JC1410200), Shanghai Jiao Tong University 2030 Initiative, and State Key Laboratory of Dark Matter Physics. We acknowledge support from the Hongwen Foundation in Hong Kong and the New Cornerstone Science Foundation in China. The computations were performed on the $\pi$ 2.0 cluster at the Center for High Performance Computing, Shanghai Jiao Tong University. We thank Professors Weiping Zhang and Guzhi Bao of Shanghai Jiao Tong University for insightful discussions about the seed photon generation and detection.

\clearpage
\bibliography{axion_wakefield_generation_prx}

\end{document}